\documentclass[prl,twocolumn,showpacs,superscriptaddress,preprintnumbers,amssymb]{revtex4}
\usepackage{graphicx}
\usepackage{dcolumn}
\usepackage{bm}
\usepackage{wasysym}

\newcommand{\beq}{\begin{equation}}
\newcommand{\eeq}{\end{equation}}
\newcommand{\beqn}{\begin{eqnarray}}
\newcommand{\eeqn}{\end{eqnarray}}

\begin{document}
\title{Quantum critical points of Helical Fermi Liquids}
\author{Cenke Xu}
\affiliation{Department of Physics, Harvard University, Cambridge,
MA 02138}
\date{\today}

\begin{abstract}

Following our previous work, we study the quantum phase
transitions which spontaneously develop ferromagnetic spin order
in helical fermi liquids which breaks continuous spin-space
rotation symmetry, with application to the edge states of 3d
topological band insulators. With finite fermi surface, the
critical point has both $z = 3$ over-damped and $z = 2$
propagating quantum critical modes, and the $z = 3$ mode will lead
to non-fermi liquid behavior on the entire fermi surface. In the
ordered phase, the Goldstone mode is over-damped unless it
propagates along special directions, and quasiparticle is ill
defined on most parts of the fermi surface except for special
points. Generalizations of our results to other systems with
spin-orbit couplings are also discussed.

\end{abstract}
 \maketitle

\begin{figure}
\includegraphics[width=3.1in]{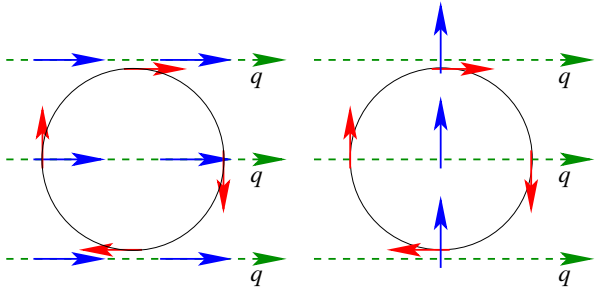}
\caption{The fermi surface of helical FL, with finite chemical
potential. The green dashed arrow is the direction of momentum
$\vec{q}$ carried by $\vec{\phi}_{\vec{q}}$, the blue arrow is the
direction of $\vec{\phi}$, the red arrow represents the helical
spin direction on the fermi surface. (Left), a longitudinal mode
of $\vec{\phi}$ with $\vec{\phi}$ parallel to $\vec{q}$ interacts
strongly with the helical fermi liquid close to $\vec{K}_f \perp
\vec{q}$, where the spin is parallel to $\vec{\phi}$; (Right), a
transverse mode of $\vec{\phi}$ interacts weakly with helical
fermi liquid.} \label{trans}
\end{figure}

Helical fermi liquids (FL) have momentum dependent spin or
pseudospin alignments at the fermi surface. For instance, the
well-studied Rashba model \cite{rashba1960,rashba1984} \beqn H =
\frac{k^2}{2m} + \alpha (k_x\sigma^y - k_y\sigma^x) \label{rashba}
\eeqn have inner and outer fermi surfaces with opposite inplane
helical spin direction. Another example of helical FL is the edge
states of 3d topological band insulators (TBI) like
$\mathrm{Bi}_{2 - x}\mathrm{Sn}_x \mathrm{Te}_3$
\cite{hasan2009a,hasan2009b,fang2008,zxshen2009,fu2007,fu2008},
which can be described by the following Dirac fermion Hamiltonian
\beqn H = v_f(k_x\sigma^y - k_y\sigma^x) \label{dirac}\eeqn $v_f$
is the fermi velocity at the Dirac point. When the chemical
potential is nonzero, the spin $\sigma^a$ of the electrons are
perpendicular with their momenta at the fermi surface (Fig.
\ref{trans}). Eq. \ref{dirac} is the minimal model of helical FL
because it has only one fermi surface, and the time-reversal
partner of this model is located on the opposite edge of the three
dimensional TBI. Recent ARPES measurement \cite{hasan2009b} has
successfully observed the helical spin alignment of the edge
states of 3d TBI. Both the Rashba model and Eq. \ref{dirac} are
invariant under the following symmetry transformations: \beqn
\mathrm{T} &:& \  t \rightarrow -t, \ k_i \rightarrow -k_i, \
\sigma^a \rightarrow - \sigma^a, \cr\cr \mathrm{P}_x &:& x
\rightarrow -x, \ \sigma^x \rightarrow \sigma^x, \ \sigma^y
\rightarrow -\sigma^y, \cr\cr \mathrm{P}_y &:& y \rightarrow -y, \
\sigma^y \rightarrow \sigma^y, \ \sigma^x \rightarrow -\sigma^x,
\cr\cr R_{\theta} &:& \ (x, y)^t \rightarrow e^{i\theta \tau^2}
(x, y)^t, \ \sigma^a \rightarrow e^{-
i\frac{\theta}{2}\sigma^z}\sigma^a e^{ i\frac{\theta}{2}\sigma^z}.
\label{symmetry}\eeqn $\mathrm{T}$, $\mathrm{P}_a$ are discrete
symmetry transformation, while $R_{\theta}$ continuously rotate
spin and space by the same angle $\theta$, which corresponds to
the conservation of total angular momentum. In this work we will
take the edge states of the TBI as an example of helical FL, but
our results can be straightforwardly generalized to other
situations.

Just like the ordinary fermi liquid, strong enough interaction can
lead to various types of instabilities of helical FL, which
spontaneously break all or part of the symmetries listed in Eq.
\ref{symmetry}. Supposedly strong interaction will play an
important role in the TBI with transition metal elements, where
the interplay between spin-orbit coupling and interaction can lead
to many interesting phenomena \cite{balents2009,qi2009}. According
to the standard Hertz-Millis theory \cite{hertz1976,millis1993},
for ordinary fermi liquid, the quantum critical modes are usually
over-damped due to low energy particle-hole excitations, which
lead to nonrelativistic universality class. In a recent paper we
studied the discrete time reversal symmetry breaking of the
helical FL \cite{cenke2009}, and the helical spin alignment at the
fermi surface strongly suppresses the coupling between order
parameter and the particle-hole excitations. Therefore the
T-breaking phase transition belongs to the $z = 1$ 3d Ising
universality class. In the current paper we will study the phase
transition that breaks the continuous symmetry $R_\theta$ in Eq.
\ref{symmetry}, which is associated with the inplane ferromagnetic
spin order $\vec{\phi} = (\phi_x, \phi_y)$. Identifying the
leading spin order instability of the helical FL requires the
detailed knowledge of the fermion interaction. As in our previous
work, we will focus on the universal physics at the quantum
critical point, assuming the existence of the phase transition.

Without loss of generality, the Lagrangian describing this
transition can be written as \beqn L &=& L_f + L_b + L_{bf},
\cr\cr L_f &=& \psi^\dagger ((i\partial_t - \mu) + i
v_f\hat{z}\cdot (\vec{\sigma} \times \vec{\nabla}))\psi, \cr\cr
L_b &=& |\partial_t \vec{\phi}|^2 - \sum_{i = x,y} v_b^2
|\partial_i\vec{\phi}|^2  - r |\vec{\phi}|^2 - u|\vec{\phi}|^4,
\cr L_{bf} &=& g \vec{\phi}\cdot \psi^\dagger \vec{\sigma}\psi.
\label{lag}\eeqn The order parameter $\vec{\phi}$ couples with the
Dirac current, which is similar to the 3d QED with gauge field
$a_\mu = (\phi_y, - \phi_x, 0)$ and the temporal gauge choice $a_0
= 0$. The temporal component $a_0$ represents the charge density
mode, which will couple to the XY spin mode after integrating out
fermions. The implication of the spin-charge coupling has been
studied in Ref. \cite{bernevig2004,sri2009}. However, at the
frequency and momentum range we are interested in, the charge mode
will not lead to singular corrections to the spin response
function, $i.e.$ the charge mode is not critical. The transition
occurs when $r = 0$. Let us take $\mu = 0$ first. $L_b$ alone
describes a 3d XY transition, but $g$ is obviously relevant at the
3d XY and free Dirac fermion fixed point based on the well-known
scaling dimensions $[\psi] = 1$ and $[\vec{\phi}] = 0.519$
\cite{hasenbusch2001}, this fermion-boson coupling will modify the
nature of this transition. A controlled starting point for the
calculation of critical exponents, is to increase the number of
fermion components to $N > 1$, and take the large-$N$ limit. After
integrating out the fermions, the renormalized boson Lagrangian
reads \beqn L_b &\sim& N\tilde{P}_{ab}\sqrt{\omega^2 +
v_f^2q^2}\phi_{a,\omega,\vec{q}}\phi_{b,-\omega,-\vec{q}} + \cdots
\cr\cr \tilde{P}_{ab} &=& \frac{\omega^2\delta_{ab} + v_f^2
q_aq_b} {\omega^2+v_f^2q^2}, \label{muzero}\eeqn
In the large-$N$ limit the scaling dimension of $\vec{\phi}$ is
$[\vec{\phi}] = 1$. A standard $1/N$ expansion calculation can be
applied to the case with large but finite $N$, although when $N =
1$ there is no small parameter for expansion. Phase transitions
with order parameters coupled to certain component of Dirac
current were studied in the context of $d-$wave superconductor by
$\epsilon$ expansion \cite{vojta2000,vojta2000b} and $1/N$
expansion \cite{kim2008,huh2008}, and a fixed point with extreme
anisotropic fermi velocity was found if we start with an
anisotropic initial condition \cite{kim2008,huh2008}.

Now let us consider the situation with $\mu \neq 0$, $i.e.$ the
situation with finite fermi surface. In the ordinary fermi liquid,
an order parameter with small momentum $|\vec{q}| \ll k_f$
interacts most strongly with fermions at $\vec{K}_f \perp
\vec{q}$, because there the particle-hole excitation at momentum
$\vec{q}$ is softest, and usually leads to over-damping of the
quantum critical modes. In our current case, since the spin
alignment at the fermi surface is determined by its momentum, not
all quantum critical modes have strong interactions with the
fermions. For instance, for a quantum critical mode with
$\vec{\phi}$ and momentum $\vec{q}$ both parallel to $\hat{x}$, it
couples with the fermions at two points $\vec{K}_f = (0, \pm k_f)$
in the same way as the ordinary fermi liquid, therefore the
longitudinal mode of $\vec{\phi}$ is over-damped (Fig.
\ref{trans}). If $\vec{\phi}$ is parallel with $\hat{y}$ while
$\vec{q}$ parallel with $\hat{x}$, since the matrix element
$\langle \psi_{\vec{k}}|\sigma^y| \psi_{\vec{k}}\rangle = 0$ when
$\vec{k} = (0, \pm k_f)$, the transverse mode of $\vec{\phi}$
should couple weakly with the fermions. These observations suggest
that after integrating out the fermions, the transverse and
longitudinal modes of $\vec{\phi}$ will behave differently.
Indeed, the $L_f + L_{bf}$ part of the Lagrangian Eq. \ref{lag} is
invariant under gauge transformation \beqn \phi_a \rightarrow
\phi_a + \epsilon_{ab}\partial_b \theta, \ \ \psi \rightarrow
e^{i\theta}\psi, \label{gauge} \eeqn and $\theta$ is an arbitrary
function of space. If we integrate out the fermions, and consider
a Feynman diagram without boson internal line, this gauge symmetry
implies that any external boson line of this diagram only involves
the longitudinal mode of $\vec{\phi}$ when the frequency of this
external line is zero. For instance, Eq. \ref{muzero} is
consistent with this conclusion, because $\tilde{P}_{ab}
\epsilon_{bc}q_c = 0$ when $\omega = 0$.

\begin{figure}
\includegraphics[width=3.3in]{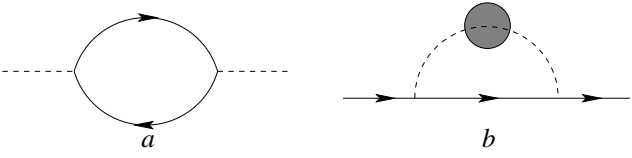}
\caption{$a$, the bubble diagram for the self-energy correction to
$\vec{\phi}$; $b$, the self-energy correction to fermions.}
\label{fd3}
\end{figure}

The above observation becomes explicit in the bubble diagram in
Fig. \ref{fd3}$a$, which renormalizes the Gaussian part of $L_b$
as \beqn \Delta L_{b} \sim \phi_{a, - \omega, - \vec{q}}
\chi(\omega, \vec{q})_{ab} \phi_{b, \omega, \vec{q}} . \eeqn If we
fix chemical potential $\mu$ and the energy at the ultraviolet
cut-off, the calculation suggests that the static and uniform
susceptibility $\chi(0,0)$ vanishes, which can be naturally
expected because a uniform order of $\vec{\phi}$ merely moves the
entire Dirac cone away from the original position, without
developing any extra polarization on the fermi surface. This
calculation implies that the mass gaps of transverse and
longitudinal modes are still equal after coupling to the fermions,
because a different mass gaps for these two modes will lead to
very singular long range interaction between $\vec{\phi}$ in real
space-time. In the ordinary Hert-Millis theory
\cite{hertz1976,millis1993} of quantum phase transition inside
fermi liquid, the most singular correction to the effective
Lagrangian of order parameter comes from the imaginary part of the
susceptibility, which corresponds to the damping of the critical
mode of order parameter $\vec{\phi}$ through particle-hole
excitations. The damping rate can be calculated from the Feynman
diagram Fig. \ref{fd3}$b$, or through the Fermi-Golden rule \beqn
\mathrm{Im}[\Sigma_{\vec{\phi}}(\omega, q)_{ab}] &\sim& \int
\frac{d^2k}{(2\pi)^2} [f(\epsilon_{k+q}) - f(\epsilon_{k})]\cr\cr
&\times& g^2 \delta(|\omega| - \epsilon_{k+q} + \epsilon_k)
\mathcal{M}^{a}_{k, k+q}\mathcal{M}^{b}_{k+q, k} \cr\cr &\sim& \
g^2 \frac{|\omega|}{ v_f q} P_{ab} + g^2 \frac{|\omega|^3}{ v_f^3
q^3} (\delta_{ab} - P_{ab}), \cr\cr M^a_{k, k+q} &=& \langle k|
\psi^\dagger_k\sigma^a\psi_{k+q} |k+q \rangle, \cr\cr P_{ab} &=&
\frac{q_aq_b}{q^2}. \label{damp}\eeqn Matrix $P_{ab}$ projects
$\vec{\phi}$ to its longitudinal part, and $P$ satisfies the
algebraic relation $P^2 = P$. This calculation indicates that only
the longitudinal part of $\vec{\phi}$ is over-damped, while the
transverse part of $\vec{\phi}$ gains a much weaker damping as
$\omega/(v_f q) \rightarrow 0$, which is consistent with our
observation. The decomposition between transverse and longitudinal
modes also occurs in a very different problem: the nematic
transition of fermi liquid \cite{kivelson2001}.

The real part of self-energy can be calculated accordingly. The
result in the infrared limit depends on how we take the limit of
the frequency and momentum. In the limit $\omega/(v_fq) \ll 1$,
the result reads \beqn \mathrm{Re}[\Sigma_{\vec{\phi}}(\omega,
q)_{ab}] &\sim& \int \frac{d^2k}{(2\pi)^2}
\frac{[f(\epsilon_{k+q}) - f(\epsilon_{k})](\epsilon_{k+q} -
\epsilon_{k})}{\omega^2 - (\epsilon_{k+q} - \epsilon_{k})^2}
\cr\cr &\times& g^2 \mathcal{M}^{a}_{k, k+q}\mathcal{M}^{b}_{k+q,
k} \cr\cr &\sim&  g^2 \int d\theta
\frac{F(\theta)_{ab}}{\frac{\omega^2}{v_f^2q^2} -
[\cos(\theta)]^2} \cr\cr &\sim& - g^2 \frac{\omega^2}{q^2}P_{ab} +
g^2 \frac{\omega^2}{q^2}(\delta_{ab} - P_{ab}) + \cdots \eeqn
$F(\theta)_{ab}$ is a spin dependent function of $\theta$ (the
angle between $\vec{k}$ and $\vec{q}$). The ellipses include less
singular terms. The transverse mode and longitudinal mode both
acquire singular correction $\omega^2/q^2$. The $\omega^2/q^2$
behavior is the reason of the existence of various collective
modes of the ordinary fermi liquid, such as the zero sound. In our
previous paper about the Ising transition \cite{cenke2009}, the
$\omega^2/q^2$ term does not show up because the matrix element
$|\langle k |\sigma^z| k+q \rangle|^2 \sim q^2$ with small $q$,
which cancels the $q^2$ in the denominator. Keeping all the
relevant terms from imaginary and real parts, the full Gaussian
Lagrangian for $\vec{\phi}$ in the Euclidean space-time reads
\beqn L_b &\sim& (Ag^2 \frac{|\omega|}{v_fq} +
v_l^2q^2)P_{ab}\phi_{a, -\omega, -\vec{q}}\phi_{b, \omega,
\vec{q}} \cr\cr  &+& (B g^2 \frac{\omega^2}{v_f^2q^2} +
v_t^2q^2)(\delta_{ab} - P_{ab})\phi_{a, -\omega, -\vec{q}}\phi_{b,
\omega, \vec{q}}. \label{gaussian}\eeqn $A$ and $B$ are order one
dimensionless constants, $v_l$ and $v_t$ are renormalized boson
velocities. The renormalized propagator of $\vec{\phi}$ reads
\beqn D(\omega, \vec{q})_{ab} \sim \frac{P_{ab}}{Ag^2
\frac{|\omega|}{v_fq} + v_l^2q^2} + \frac{\delta_{ab} -
P_{ab}}{Bg^2 \frac{\omega^2}{v_f^2q^2} + v_t^2q^2}.
\label{prop}\eeqn This calculation suggests that by coupling to
the helical FL, at the quantum critical point the longitudinal
part of $\vec{\phi}$ becomes a $z = 3$ over-damped mode, while the
transverse part of $\vec{\phi}$ becomes a $z = 2$ propagating
mode. For both $z = 3$ and $z = 2$ scaling, $\omega/q \rightarrow
0$ with small frequency and momentum, which justifies the small
$\omega/(v_fq)$ limit we took at the beginning of our calculation.
Because $\chi(0,0)$ vanishes, both transverse and longitudinal
modes become gapless at the same critical point.

In two dimension, $z = 2$ quantum critical point is at the upper
critical dimension, therefore the Gaussian fixed point Eq.
\ref{gaussian} is stable with marginally irrelevant perturbations.
As discussed in ordinary $z = 3$ and $z = 2$ quantum critical
points, more singular perturbations may be generated with higher
order fermion loop diagrams \cite{vojta1997,chubukov2004}. In
principle the stability of the one-loop result requires either
careful analysis of the higher loop expansions or nonperturbative
approach as Ref. \cite{lawler2006}. We will discuss this in
future, right now we tentatively focus on the one-loop result. The
$z = 3$ and $z = 2$ quantum critical modes define the quantum
critical regime $T
> |r|^{z\nu} \sim |r|^{3/2}$ and $T
> |r|$ respectively, therefore in the quantum critical regime
at low temperature the thermal dynamics is dominated by the $z =
3$ longitudinal mode, which leads to the specific heat scaling
$C_v \sim T^{2/3}$.

In the quantum critical regime, the fermions interact strongly
with the bosons, and gain self-energy renormalization. The
self-energy renormalization can be calculated in the standard way
using diagram Fig. \ref{fd3}$b$. Let us take the quasiparticle at
$\vec{K}_f = (+k_f, 0) $, where the dispersion of quasiparticle
can be expanded as $\epsilon_k \sim v_fk_x + v_y k_y^2$, therefore
around this point $k_x$ has scaling dimension 2 while $k_y$ has
scaling dimension 1. Since the spin is along the $\hat{y}$
direction, in the calculation we should project the boson
propagator in the $\hat{y}$ direction. Using the propagator in Eq.
\ref{prop}, the fermion self-energy reads \beqn \Sigma_{\psi}
&\sim& \int
\frac{d^2kd\nu}{(2\pi)^3} \frac{1}{i(\omega + \nu) - v_f (k_x +
q_x) - v_y^2(k_y + q_y)^2} \cr\cr &\times& (\frac{P_{yy}}{Ag^2
\frac{|\omega|}{v_fk} + v_l^2k^2} + \frac{1 -
P_{yy}}{Bg^2\frac{\omega^2}{v_f^2k^2} + v_t^2k^2}).
\label{self}\eeqn
The self-energy is dominated by the $z = 3$ longitudinal mode,
which leads to the same self-energy scaling as the ordinary $z =
3$ quantum critical point: \beqn \Sigma_{\psi}(\omega)'' \sim
|\omega|^{2/3}\mathrm{sgn}[\omega]. \label{nonfermi}\eeqn
Therefore at the quantum critical point the system has non-fermi
liquid behavior.

It is useful to discuss more about the isolated fermi patch around
$\vec{K}_f = (+k_f, 0)$. As in Eq. \ref{self}, after projecting
the boson propagator along $\hat{y}$ direction, the transverse
part of the propagator acquires a factor $1 - P_{yy} =
k_x^2/(k_x^2 + k_y^2)$. Because now $k_x$ has scaling dimension 2,
the transverse part of the propagator is effectively suppressed.
Therefore in the Lagrangian we can keep just the scalar
longitudinal mode. Now the isolated patch is described by the
following scaling invariant Lagrangian \beqn L &=& L_f + L_b +
L_{bf}, \cr\cr L_f &=& \sum_{\omega,
\vec{k}}(c|\omega|^{2/3}\mathrm{sgn}[\omega] - v_f k_x - v_y
k_y^2)\psi^\dagger_{\omega, \vec{k}}\psi_{\omega, \vec{k}}, \cr\cr
L_b &=& \sum_{\omega, \vec{k}}(Ag^2 \frac{|\omega|}{v_f|k_y|} +
v_l^2k^2_y)\phi_{\omega, \vec{k}} \phi_{-\omega, -\vec{k}} ,
\cr\cr L_{bf} &=&\sum_{\omega, \nu, \vec{p}, \vec{q}}
g\phi_{\omega, \vec{q}}\psi^\dagger_{\nu+\omega,
\vec{p}+\vec{q}}\psi_{\vec{q}}. \eeqn One can verify that under
the scaling transformation $\omega \rightarrow \omega b^3$, $k_x
\rightarrow k_x b^2$, $k_y \rightarrow k_y b$, $\psi \rightarrow
\psi b^4$, $\phi \rightarrow \phi b^4$, $g \rightarrow g$ this
Lagrangian is invariant. Therefore the coupling $g$ is a marginal
perturbation, and the theory becomes identical to the spinon and
gauge field problem discussed in Ref.
\cite{sslee2009,polchinksi1994}. In the large-$N$ limit with $N$
copies of the fermi patches, this theory is expected to be
controlled by a strongly coupled CFT \cite{sslee2009}.

Now let us discuss the ordered phase, with $r < 0$ in Eq.
\ref{lag}. The low energy physics of the ordered phase is
dominated by the Goldstone mode. Let us assume the order is
$\langle \vec{\phi} \rangle \sim (0, \phi_y)$, and the Goldstone
mode is $\phi_x$. As already mentioned before, in the ordered
phase, the entire Dirac cone is translated in the momentum space
without change of the shape of fermi surface, therefore the
Goldstone mode of ordered phase of $\vec{\phi}$ has a very similar
behavior as the quantum critical mode: \beqn L_{\phi_x} \sim
Ag^2\frac{i |\omega| q_x^2}{v_fq^3} +
Bg^2\frac{\omega^2}{v_f^2q^2}(\frac{q_y^2 - q_x^2}{q^2} +
D\frac{q_x^2q_y^2}{|r| q^4}) - v^2 q^2 .\eeqn When the momentum is
along $\hat{x}$ (longitudinal), this Goldstone mode is an
over-damped $z = 3$ mode; when momentum is along $\hat{y}$
(transverse), the Goldstone mode is a propagating $z = 2$ mode. In
the ordered phase, the over-damped longitudinal Goldstone mode
will lead to the same nonfermi liquid self-energy correction as
Eq. \ref{nonfermi} for almost all points on the fermi surface:
\beqn \Sigma_\psi(\omega)'' \sim
|\cos(\phi)|^{4/3}|\omega|^{2/3}\mathrm{sgn}[\omega],\eeqn $\phi$
is the angle between $\vec{K}_f$ and $\vec{\phi}$. At the special
points $\phi = \pm \pi/2$, the correction to the fermion
self-energy mainly comes from the transverse part of the Goldstone
mode, which reads \beqn \Sigma_\psi(\omega)'' \sim
|\omega|^{3/2}\mathrm{sgn}[\omega]. \eeqn Therefore in the ordered
phase only two special points of the fermi surface have
well-defined quasiparticles. The critical dynamics and the fermion
self energy behavior in the ordered XY phase are similar to the
nematic transition in 2d fermi liquid \cite{kivelson2001}.

So far we only kept the lowest order momentum dependent terms in
Eq. \ref{dirac}, while in real system the symmetry $R_{\theta}$ is
broken by lattice symmetry, as was shown by first principle
calculations \cite{liu2009}. In $\mathrm{Bi}_{2 - x}\mathrm{Sn}_x
\mathrm{Te}_3$ the continuous O(2) symmetry of $R_{\theta}$ is
broken down to $C_6$ inplane rotation symmetry with large chemical
potential \cite{zxshen2009}, and the spin will be canted along $z$
direction except for isolated points on the fermi surface
\cite{liu2009,fu2009}. In our previous work we have argued that
the $z$ direction canting will lead to damping of the Ising order
parameter $\phi \sim \psi^\dagger \sigma^z \psi$. The $XY$ order
parameter will still be decomposed into damped part and undamped
part, although both parts will only have discrete rotation
symmetry.

We have used the edge states of 3d TBI as the example of helical
FL, our analysis is applicable to other helical FL. For instance,
if we give graphene a small but finite chemical potential, we can
consider the spontaneous generation of order
$\bar{\psi}\vec{\gamma}\mathcal{T}^a\psi$. $\mathcal{T}^a \in
\mathrm{SU}(4)$ is the flavor symmetry matrix operating on the
real spin and Dirac cone valley space. The gamma matrices
$\gamma_i$ which operate on the two sublattices plays the role as
the helical spin in our analysis, and the results in our paper are
still applicable. Our study can also be applied to the Rashba
model in Eq. \ref{rashba}, as long as the momentum $\vec{q}$
carried by the boson field $\vec{\phi}_{\vec{q}}$ is much smaller
than the distance between the two fermi surfaces of Rashba model:
$\vec{q} \ll |\vec{K}_{f, out}| - |\vec{K}_{f, in}|$, therefore
$\vec{\phi}$ does not induce coupling between the two fermi
surfaces. One important difference between the Rashba model and
the edge states of TBI is that, the uniform and static
susceptibility is nonzero for Rashba model, although the
transverse and longitudinal modes still gain the same uniform and
static correction.

Generalization of these results to other spin-orbit coupled
electron models is straightforward. The most general form of
spin-orbit coupled electron model is \beqn H = \frac{k^2}{2m} +
\sum_a \beta B(\vec{k})_a\cdot\sigma_a. \eeqn Here the two bands
of the Pauli matrices can stand for different physical degrees of
freedom depending on the context, for instance this Hamiltonian
can also represent hybridization between different orbital states.
When $B(\vec{k})_a$ takes the $p$-wave form $B_a \sim
\epsilon_{ab}k_b$, the model becomes the Rashba model. For
example, let us consider the following model as a 3d version of
Rashba model: $B_a \sim k_a$, $a = x, y, z$. This model also has
inner and outer fermi surfaces with hedgehog spin distribution
around the fermi surface. The system has O(3) rotation symmetry as
long as spin and space rotation are synchronized. It would be
interesting to consider the spontaneous breaking of this
continuous symmetry by developing nonzero order of $\vec{\phi} =
(\phi_x, \phi_y, \phi_z)$. After integrating out the fermions,
just like the two dimensional cases we considered above, the
vector $\vec{\phi}$ is decomposed into one longitudinal mode and
two transverse modes. At the quantum critical point, the
longitudinal mode is a $z = 2$ propagating mode while the two
transverse modes are $z = 3$ over-damped modes. We can also
consider the possibility of $d$-wave $B(\vec{k})_a$, for instance
$B_x \sim k_x^2 - k_y^2$, $B_y \sim 2k_xk_y$, which can be
realized in hole-doped GaAs quantum well with inversion symmetric
confining potential \cite{qi2006}. The result of our paper is
still applicable to this model as long as we define the projection
matrix in Eq. \ref{gaussian} and Eq. \ref{prop} as $P_{ab} \sim
B_aB_b/(B_a^2 + B_b^2)$.

As a summary, in this work we discussed the spontaneous breaking
of the continuous spin-space combined rotation symmetry in helical
fermi liquid. we calculated the quantum critical modes at the
quantum critical point, and the Goldstone mode in the ordered
phase, as well sa their effects on the quasiparticles. However, as
was mentioned already, higher order loop diagrams may lead to more
singular momentum and frequency dependent terms which have the
potential to destroy the Gaussian fixed point studied in our
paper. Therefore the analysis of the Gaussian fixed point in this
paper is the basis of our future studies.



\bibliography{domain}

\end{document}